\documentclass{article}
\usepackage{amsmath}
\usepackage{amssymb}
\usepackage{amsfonts}
\usepackage{graphicx} 
\begin{document}

\title{On analytical solution of stationary two dimensional boundary problem of
natural convection}
\author{Sergiej Leble and Witold M.Lewandowski*\\Gda\'{n}sk University of Technology\\Department of Differential Equations and Applied Mathematics\\*Department of Chemical Apparatus and Machinery}
\maketitle

\begin{abstract}
Approximate analytical solution of two dimensional problem for stationary
Navier-Stokes, continuity and Fourier-Kirchhoff equations describing free
convective heat transfer from isothermal surface of half infinite vertical
plate is presented. The problem formulation is based on the typical for
natural convection assumptions: the fluid noncompressibility and Boussinesq
approximation. We also assume that orthogonal to the plate component of
velocity is small. Apart from the basic equations it includes
boundary conditions: the constant temperature and zero velocity on the plate.
At the starting point of the flow we fix average temperature and vertical
component of velocity, as well as basic conservation laws in integral form.
The solution of the boundary problem is represented as a Taylor Series in
horizontal variable with coefficients depending on vertical variable.
\end{abstract}

\section{ Introduction}

The results of theoretical and experimental study of free convective flows
from heating objects are widely published and they are very useful to
determine convective heat losses from apparatus, devices, pipes in industrial
or energetic installations, electronic equipment, architectonic objects and so
on by engineers and designers.

The problem of convective flow development traditionally is based either on
boundary layer theory or on self-similarity theory \cite{1}, \cite{2},
\cite{3}. Both methods of natural convection heat transfer description use
simplifications that allow to transform the basic fundamental equations of
Navier-Stokes, mass and Fourier-Kirchhoff \ equations. As a main point of the
methods is an ordinary differential equation introduction which solution
relates to a general problem by some contraction procedure. Such description
don't give a possibility to investigate some details of a flow field for
example at vicinity of its starting point. It is known that the starting point
influence is important in many aspects of fluid flow development, especially
in critical conditions.

There are some problems of convection for which the boundary layer approach
fails. One of such cases is an assumption that boundary layer starts on a
plate, that contradicts experiment. The other is conceptual for the theory:
the fluid flow is restricted by a conditional boundary which also don't exist
in reality. Due to this Prandtl introduced integration across a boundary layer
to determine average values of fluid flow (velocity, temperature).

The similarity solution method is based on a specific combination of
independent variables introduction that allows to get rid of
two-dimensionality of the general problem. The resulting ordinary differential
equation gives an information of velocity and temperature profiles that is an
advance in comparison with boundary layer theory. However such description
still is rather qualitative than quantitative. For example it doesn't give an
answer about flow behavior in the vicinity of the starting point in the flow.
We would stress that an interest to the problem still exists \cite{4},
\cite{5}, \cite{6}.

We consider approximate analytical solution of stationary convective fluid
flow induced by an isothermal vertical surface with the axis $y$ parallel to
gravity force $x$ is used as a horizontal one. The choice of the coordinate
system is typical for laminar natural convection simplifications with the
following main assumptions: fluid is incompressible, Boussinesq approximation,
normal to the surface component of velocity is neglected \cite{3,7}. The case
of isothermal vertical surface is obtained as a limit of horizontal conic with
a base angle of the conic $\alpha=0$ \cite{8,9}

Our solutions are build without use neither of boundary layer nor self -
similarity concepts. We are looking for the solution as Taylor series in $x$
and derive a system of equations for coefficient of the expansion. Each
cutting of the series results in polynomial approximation in $x$ and find a
closed system of equations in $y$. We restrict ourselves by upper half plane
($y\geqslant0)$. Such approach needs a formulation of boundary conditions in
the starting point of the flow. We take into account the physical character of
the flow: namely its natural convection origin that means average velocity is
zero at $y=0$ level. For the formulation of boundary conditions we base on
conservation laws in integral form. 

\section{The basic equations}

Let us consider a two dimensional stationary flow of incompressible fluid in
the gravity field. The flow is generated by a convective heat transfer from
solid plate to the fluid. The plate is isothermal and vertical. In the
Cartesian coordinates $x,y$ the Navier-Stokes (NS) system of equations have
the form \cite{1}:
\begin{equation}
\rho\left(  W_{x}\frac{\partial W_{y}}{\partial x}+W_{y}\frac{\partial W_{y} 
}{\partial y}\right)  =g\rho_{\infty}b\left(  T-T_{\infty}\right)
-\frac{\partial p}{\partial y}+\rho\nu\left(  \frac{\partial^{2}W_{y} 
}{\partial y^{2}}+\frac{\partial^{2}W_{y}}{\partial x^{2}}\right)  \label{NSy} 
\end{equation} 
\begin{equation}
\rho\left(  W_{x}\frac{\partial W_{x}}{\partial x}+W_{y}\frac{\partial W_{x} 
}{\partial y}\right)  =-\frac{\partial p}{\partial x}+\rho\nu\left(
\frac{\partial^{2}W_{x}}{\partial y^{2}}+\frac{\partial^{2}W_{x}}{\partial
x^{2}}\right)  \label{NSx} 
\end{equation}
In the above equations the pressure terms are divided in two parts
$\widetilde{p}=p_{0}+p$. The first of them is the hydrostatic one that is
equal to mass force $-g\rho_{\infty}$, where $\rho=\rho_{\infty}\left(
1-b\left(  T-T_{\infty}\right)  \right)  =\rho_{\infty}\rho^{\prime}$ is the
density of \ a liquid at the nondisturbed area where the temperature is
$T_{\infty}$. The second one is the extra pressure denoted by $-\nabla p.$The
part of gravity force $gb\left(  T-T_{\infty}\right)  $ arises from dependence
of the extra density on temperature, $b$ is a coefficient of thermal expansion
of the fluid. In the case of gases \bigskip$b=-\frac{1}{\rho}\left(
\frac{\partial\rho}{\partial T}\right)  _{p}=\frac{1}{T_{\infty}}.$The last
terms of the above equations represents the friction forces with the kinematic
coefficient of viscosity $\nu.$

The mass continuity equation in the conditions of natural convection of
incompressible fluid in the steady state \cite{2} has the form:. 
\begin{equation}
\frac{\partial W_{x}}{\partial x}+\frac{\partial W_{y}}{\partial y}=0.
\label{div} 
\end{equation}
The temperature dynamics is described by the stationary Fourier-Kirchhoff (FK) equation: 
\begin{equation}
W_{x}\frac{\partial T}{\partial x}+W_{y}\frac{\partial T}{\partial y}=a\left(
\frac{\partial^{2}T}{\partial y^{2}}+\frac{\partial^{2}T}{\partial x^{2} 
}\right)  \label{FK} 
\end{equation}
where $W_{x}$ and $W_{y}$ are the components of the fluid velocity
$\overline{W}$ that are shown on the Fig.1; $T$, $p$ - temperature and
pressure disturbances correspondingly and $a$ is the thermal
diffusivity. 
\begin{figure}
[ptb]
\begin{center}
\includegraphics[
height=3.4748in,
width=4.4547in
] 
{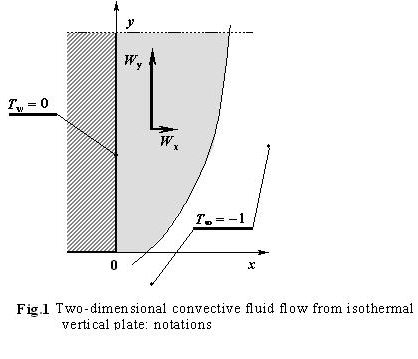} 
\end{center}
\end{figure}
From the point of clarity of further transformations we use the same scale $l$
along both variables $x$ and $y$. We will return to the eventual difference
between characteristic scales in different directions while the solution
analysis to be provided. After introducing nontraditional variables:
\[
x^{\prime}=x/l,y^{\prime}=y/l,T^{\prime}=(T-T_{w})/(T_{w}-T_{\infty}),
\] 
\begin{equation}
p^{\prime}=p/p_{\infty},W_{x}^{\prime}=W_{x}/W_{o},W_{y}^{\prime}=W_{y}/W_{o}
\label{x'} 
\end{equation}
we obtain in Boussinesq approximation (in all terms besides of buoyancy one we
put $\rho\thickapprox\rho_{\infty}$ ).\bigskip 
\begin{equation}
W_{x}^{\prime}\frac{\partial W_{y}^{\prime}}{\partial x^{\prime}} 
+W_{y}^{\prime}\frac{\partial W_{y}^{\prime}}{\partial y^{\prime}}=L\left(
T^{\prime}+1\right)  -\frac{p_{\infty}}{\rho_{\infty}W_{o}^{2}}\frac{\partial
p^{\prime}}{\partial y^{\prime}}+\nu^{\prime}\left(  \frac{\partial^{2} 
W_{y}^{\prime}}{\partial y^{\prime2}}+\frac{\partial^{2}W_{y}^{\prime} 
}{\partial x^{\prime2}}\right)  \label{NS-1} 
\end{equation} 
\begin{equation}
W_{x}^{\prime}\frac{\partial W_{x}^{\prime}}{\partial x^{\prime}} 
+W_{y}^{\prime}\frac{\partial W_{x}^{\prime}}{\partial y^{\prime}} 
=-\frac{p_{\infty}}{\rho_{\infty}W_{o}^{2}}\frac{\partial p^{\prime}}{\partial
x^{\prime}}+\nu^{\prime}\left(  \frac{\partial^{2}W_{x}^{\prime}}{\partial
y^{^{\prime}2}}+\frac{\partial^{2}W_{x}^{\prime}}{\partial x^{\prime2} 
}\right)  \label{NS-2} 
\end{equation}
and FK equation is written as 
\begin{equation}
W_{x}^{\prime}\frac{\partial T^{\prime}}{\partial x^{\prime}}+W_{y}^{\prime
}\frac{\partial T^{\prime}}{\partial y^{\prime}}=a^{\prime}\left(
\frac{\partial^{2}T^{\prime}}{\partial y^{\prime2}}+\frac{\partial
^{2}T^{\prime}}{\partial x^{\prime2}}\right)  \label{FK'} 
\end{equation}
where $\frac{\nu}{lW_{o}}=\nu^{\prime},\frac{a}{lW_{o}}=a^{\prime} 
,\frac{gb(T_{w}-T_{\infty})l}{W_{o}^{2}}=L,$ $l$ is a characteristic linear
dimension and $W_{0}$ is characteristic velocity.
If
\begin{equation}
W_{o}=\frac{\nu}{l}, \label{W0} 
\end{equation}
then $a^{\prime}=\Pr$ , $\nu^{\prime}=1$ and $L=G_{r},$ where
\begin{equation}
G_{r}=\frac{gb(T_{w}-T_{\infty})l^{3}}{\nu^{2}}. \label{Gr} 
\end{equation}
(Grashof number) such link will be used in final solution.
After cross differentiation of equations ($\ref{NS-1})$ and ($\ref{NS-2}$) we have:
\[
\frac{\partial}{\partial x^{\prime}}\left[  W_{x}^{\prime}\frac{\partial
W_{y}^{\prime}}{\partial x^{\prime}}+W_{y}^{\prime}\frac{\partial
W_{y}^{\prime}}{\partial y^{\prime}}-G_{r}\left(  T^{\prime}+1\right)
-\left(  \frac{\partial^{2}W_{y}^{\prime}}{\partial y^{\prime2}} 
+\frac{\partial^{2}W_{y}^{\prime}}{\partial x^{\prime2}}\right)  \right]  =
\]
\begin{equation}
=\frac{\partial}{\partial y^{\prime}}\left[  W_{x}^{\prime}\frac{\partial
W_{x}^{\prime}}{\partial x^{\prime}}+W_{y}^{\prime}\frac{\partial
W_{x}^{\prime}}{\partial y^{\prime}}-\left(  \frac{\partial^{2}W_{x}^{\prime} 
}{\partial y^{\prime2}}+\frac{\partial^{2}W_{x}^{\prime}}{\partial x^{\prime
2}}\right)  \right]  \label{NS12} 
\end{equation}
The FK equation rescales as
\begin{equation}
\Pr\left(  W_{x}^{\prime}\frac{\partial T^{\prime}}{\partial x^{\prime}} 
+W_{y}^{\prime}\frac{\partial T^{\prime}}{\partial y^{\prime}}\right)
=\left(  \frac{\partial^{2}T^{\prime}}{\partial y^{\prime2}}+\frac
{\partial^{2}T^{\prime}}{\partial x^{\prime2}}\right)  \label{FK} 
\end{equation}
Next we would formulate the problem of free convection over the heated
vertical isothermal\ plate $x=0,$ $y\in\lbrack0,l)$, dropping the primes.
\section{Method of solution and approximations}
Next aim of this paper is the theory application to the standard example of a
finite vertical plate. In this case we assume the angle between the plate and
a stream line is small that means a possibility to neglect the horizontal
component of velocity of fluid, denoting the vertical component as $W\left(
y,x\right)  $.
It means that the number of equations: ($\ref{NS12}$), ($\ref{FK}$),
($\ref{div}$) exceeds the number of variables: $T$ and $W$, hence one of the
equations should be excluded. Having in mind mass conservation account in
integral form, we exclude the equation ($\ref{x'}$). The mass conservation law
introduces boundary conditions as a link between mass law flow on levels $y=0$
and $y=L$ (see the section on boundary conditions).
Consider the power series expansions of the velocity and temperature in
Cartesian coordinates: 
\begin{equation}
W\left(  x,y\right)  =\gamma\left(  y\right)  x+\alpha\left(  y\right)
x^{2}+\beta\left(  y\right)  x^{3}+\varkappa(y)x^{4}..., \label{wyx} 
\end{equation}
\begin{equation}
T\left(  x,y\right)  =C\left(  y\right)  x+A\left(  y\right)  x^{2}+B\left(
y\right)  x^{3}+F\left(  y\right)  x^{4}... .\label{tyx} 
\end{equation}
According to standard boundary conditions on the plate we assume that the both
functions tend to zero when $x\rightarrow0$, we choose for a calculation
convenient the zero value for nontraditional temperature ($\ref{x'}$).It means
that the value of $T(x,y)$ outside of the convective flow tends to $-1.$

We would like to restrict ourselves by the fourth order approximation for both
variables that means we neglect higher order terms starting from fifth one
(see the Fig.1 where the area of the approximations validity is marked as
dashed one).

As it will be clear from further analysis we should consider the functions:
$\alpha\left(  y\right)  ,$ $\beta\left(  y\right)  ,$ $C\left(  y\right)  $
and $B\left(  y\right)  $ as variables of the first order, while
$\gamma\left(  y\right)  $ and $F\left(  y\right)  $ to be the second one.
From the relations that appear after substitution of ($\ref{wyx})$ and
($\ref{tyx})$ into ($\ref{FK'})$ and ($\ref{NS1+2})$ it follows that $A\left(
y\right)  =0.$

Substituting expressions ($\ref{wyx})$ and ($\ref{tyx})$ into the equations
($\ref{NS1+2})$ and ($\ref{FK''})$ taking into account the assumption that
$W_{x}=0$ and $W_{y}=W$ yields 
\begin{equation}
\frac{\partial}{\partial x}\left[  W\frac{\partial W}{\partial y}-G_{r}\left(
T+1\right)  -\left(  \frac{\partial^{2}W}{\partial y^{2}}+\frac{\partial^{2} 
W}{\partial x^{2}}\right)  \right]  =0, \label{NS-a} 
\end{equation}
\begin{equation}
Pr W\frac{\partial T}{\partial y}=\left(  \frac{\partial^{2}T}{\partial
y^{2}}+\frac{\partial^{2}T}{\partial x^{2}}\right). \label{FK-a} 
\end{equation}
From ($\ref{FK-a})$ one found that $A(y)=0$ and $F\left(  y\right)  =0.$
Finally from both equations ($\ref{NS-a}),$ ($\ref{FK-a})$ we obtain the
system of equations for the coefficients $B\left(  y\right)  $, $C\left(
y\right)  $, $\alpha\left(  y\right)  $ and $\beta\left(  y\right)  :$\bigskip 
\begin{equation}
6B\left(  y\right)  +\frac{\partial^{2}C\left(  y\right)  }{\partial y\partial
y}=0, \label{BC} 
\end{equation}
\begin{equation}
\Pr\alpha\left(  y\right)  \frac{\partial C\left(  y\right)  }{\partial
y}-\frac{\partial^{2}B\left(  y\right)  }{\partial y\partial y}=0,
\label{alfaBC} 
\end{equation}
\begin{equation}
-6\beta\left(  y\right)  -G_{r}C\left(  y\right)  =0, \label{betaC} 
\end{equation}
\begin{equation}
\bigskip\gamma\left(  y\right)  \frac{\partial\gamma\left(  y\right)
}{\partial y}-\frac{\partial^{2}\alpha\left(  y\right)  }{\partial y\partial
y}=0. \label{alfa} 
\end{equation}
The first two ($\ref{BC}),$ ($\ref{alfaBC})$ arise from FK equation and the
rest of them are from the NS one. \ The system of equations is closed if
$\gamma\left(  y\right)  =const=\gamma$. It means that the number of equations
and the number of unknown functions is the same.
Finally, in the first approximation the velocity and temperature are expressed as: 
\begin{equation}
W\left(  y,x\right)  =\gamma x+\alpha\left(  y\right)  x^{2}+\beta\left(
y\right)  x^{3},\ \ \ \ \ T\left(  y,x\right)  =C\left(  y\right)  x+B\left(
y\right)  x^{3}. \label{twyx} 
\end{equation}
From ($\ref{alfa}$) one has
\begin{equation}
\alpha\left(  y\right)  =C_{1}y+C_{2}. \label{C1C2} 
\end{equation}
From ($\ref{BC}$) it follows that $B\left(  y\right)  =-\frac{1}{6} 
\frac{\partial^{2}C\left(  y\right)  }{\partial y\partial y},$hence
($\ref{alfaBC}$) goes to:\bigskip$\allowbreak$ 
\begin{equation}
\frac{1}{6}\frac{\partial^{4}C\left(  y\right)  }{\partial y\partial y\partial
y\partial y}+\Pr\left(  yC_{1}+C_{2}\right)  \frac{\partial C\left(  y\right)
}{\partial y}=0. \label{C} 
\end{equation}
The equation ($\ref{betaC}$) reads: 
\begin{subequations}
\begin{equation}
\beta\left(  y\right)  =-\frac{G_{r}}{6}C\left(  y\right)  . \label{BETA} 
\end{equation}
The equation ($\ref{C})$ is the ordinary differential equation of the fourth
order, therefore its solution needs four constants of integration, denoted as:
$C(0)$, $C\prime(0)$, $C^{\prime\prime}(0)$ and $C^{\prime\prime\prime}(0).$
These constants depend on two parameters $C_{1}$ and $C_{2}$, which enter the
coefficients of the Eq.$\ref{C}$. The function $C(y)$ defines the rest
functions $\beta\left(  y\right)  $ and $B\left(  y\right)  $ via above
relations. It means that we have six constants determining the solution of
problem and we need also six corresponding boundary conditions at $y=0.$
\section{Boundary conditions formulation}
At the starting horizontal edge of the vertical plate the bouncy force yet
does not act therefore we assume that the mean vertical component of the
velocity is negligibly small. Thus we neglect a fine velocity structure of
transition area. Similar the temperature values are averaged in the vicinity
of the boundary edge point and taken as value -1 (temperature of incoming from
the bottom flow). Then let us define the mean value of a function $F(x)$
(velocity or temperature in the interval $x\in\left[  0,1\right]  $ (Fig.2) in
the nontraditional form as: $\overline{F}=$ $ 
 {\displaystyle\int\limits_{0}^{1}}
F(x)dx$. In dimensional form the interval of averaging length is $l$ which
hence is defined by this integration boundary. Let us remind that scale $l$ is
connected with special (local, horizontal) Grashof number $G_{r}$ ($\ref{Gr}$). 
 \begin{figure}
[ptb]
\begin{center}
\includegraphics[
natheight=5.646400in,
natwidth=6.046800in,
height=3.4549in,
width=3.6988in
] 
{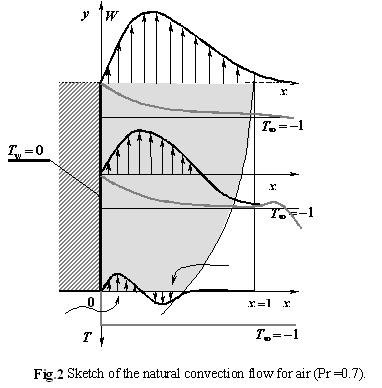} 
\end{center}
\end{figure}
 In natural convection problem theory vertical velocity component and
temperature values in surrounding of the heated plate are: $W_{y}(0,0)=0$,
$T(0.0)=-1$ according to the above notations. Taking it into account we put:
\end{subequations}
\begin{equation}
\overline{W\left(  0,x\right)  }=\left[  \
{\displaystyle\int\limits_{0}^{1}}
\left(  \gamma x+\alpha\left(  y\right)  x^{2}+\beta\left(  y\right)
\ x^{3}\right)  dx\right]  _{y\rightarrow0}=\frac{1}{2}\gamma+\frac{1} 
{3}\alpha\left(  0\right)  +\frac{1}{4}\beta\left(  0\right)  =0 \label{wsr} 
\end{equation}
therefore due to ($\ref{C1C2}$) we arrive at the first boundary condition: 
\begin{equation}
\alpha\left(  0\right)  =C_{2}=-\frac{3}{4}\beta\left(  0\right)  -\frac{3} 
{2}\gamma=\frac{L}{8\nu}C\left(  0\right)  -\frac{3}{2}\gamma\label{w-I} 
\end{equation}
Nontraditional temperature of the fluid at the lower half plane, according to
above, is $-1$. In the analogy to the condition for velocity we assume that
average temperature at the limit $y=0$ is also $-1$, therefore from
($\ref{twyx}$) $\overline{T\left(  y=0,x\right)  }= 
{\displaystyle\int\limits_{0}^{1}}
\left(  C\left(  0\right)  x+B\left(  0\right)  x^{3}\right)  dx=\allowbreak
\frac{1}{4}B\left(  0\right)  +\frac{1}{2}C\left(  0\right)  =-1.$ 
\bigskip\ Then the next boundary relation arises:
\begin{equation}
B\left(  0\right)  +2C\left(  0\right)  =-4. \label{B0C0} 
\end{equation}
Plugging $B\left(  0\right)  =-\frac{1}{6}\left[  \frac{\partial^{2}C\left(
y\right)  }{\partial y\partial y}\right]  _{y=0}=-\frac{1}{6}C^{\prime\prime
}(0)$ from ($\ref{B}$) into ($\ref{B0C0}$) gives the second boundary condition: 
\begin{equation}
\frac{1}{6}C^{\prime\prime}(0)-2C\left(  0\right)  =4. \label{war-2} 
\end{equation}
The third and forth boundary condition arize on the level $y=L$, where the
fluid lose the contact with the heated plate. At this point we suppose that 
\begin{equation}
\frac{\partial T}{\partial y}|_{y=L}=0 \label{gradT} 
\end{equation}
and
\begin{equation}
\frac{\partial W}{\partial y}|_{y=L}=0 \label{gradW} 
\end{equation}
Both conditions ($\ref{gradT}$) and ($\ref{gradW}$) have a transparent
physical meaning: the temperature and velocity does not grow from the level
$y=0$. The immediate consequence of the equation ($\ref{gradT}$) is 
\begin{equation}
\frac{\partial C\left(  y\right)  }{\partial y}|_{y=L}=0 \label{C'} 
\end{equation}
and
\begin{equation}
\frac{\partial B\left(  y\right)  }{\partial y}|_{y=L}=-\frac{1}{6} 
\frac{\partial^{3}C\left(  y\right)  }{\partial y\partial y\partial y} 
|_{y=L}=0 \label{C'''} 
\end{equation}
From ($\ref{gradW}$) one can see that 
\begin{equation}
\frac{\partial\alpha\left(  y\right)  }{\partial y}|_{y=L}=C_{1}=0
\label{alfa'} 
\end{equation}
and, according to ($\ref{BETA}$), \bigskip$\frac{\partial\beta\left(
y\right)  }{\partial y}|_{y=L}=-\frac{G_{r}}{6}\frac{\partial C\left(
y\right)  }{\partial y}|_{y=L}=0$ coincide with ($\ref{C'}$).
Looking for the rest boundary conditions let us apply conservation laws of
mass, momentum and energy for derivation of boundary conditions at $y=0$. The
first of them is the conservation of mass in steady state in two dimensions
looks as : 
\begin{equation} 
{\displaystyle\int\limits_{\Sigma}}
\rho\overrightarrow{W}\cdot\overrightarrow{n}dS=0 \label{mass} 
\end{equation}
where $\Sigma$ is a closed surface of a control volume, which cross-section is
shown in Fig.2.

The second and third of them are produced by the law of momentum vector
conservation, which is represented by Navier-Stokes equations ($\ref{NS-1})$
and ($\ref{NS-2}$).
Let us denote
$\frac{\partial p}{\partial y}=\rho_{\infty}gb\left(  T-T_{\infty}\right)
-\rho W_{x}\frac{\partial W_{y}}{\partial x}-\rho W_{y}\frac{\partial W_{y} 
}{\partial y}+\rho\nu\left(  \frac{\partial^{2}W_{y}}{\partial y^{2}} 
+\frac{\partial^{2}W_{y}}{\partial x^{2}}\right)  =E_{y\text{ }}$
and
$\frac{\partial p}{\partial x}=-\rho W_{x}\frac{\partial W_{x}}{\partial
x}-\rho W_{y}\frac{\partial W_{x}}{\partial y}+\rho\nu\left(  \frac
{\partial^{2}W_{x}}{\partial y^{2}}+\frac{\partial^{2}W_{x}}{\partial x^{2} 
}\right)  =E_{x}$, \ then one have the equation: 
\begin{equation}
\frac{\partial E_{y}}{\partial x}-\frac{\partial E_{x}}{\partial y} 
=rot_{z}\overrightarrow{E}=0 \label{rotE0} 
\end{equation}
that we use as a basic one. The integral version of this equation is obtained
by Stokes theorem: 
\begin{equation} 
{\displaystyle\int\limits_{S}}
rot_{n}\overrightarrow{E}dS= 
{\displaystyle\int\limits_{s}}
\overrightarrow{E}d\overrightarrow{s}=0, \label{rotE} 
\end{equation}
where $s$ is the boundary curve of the integration area $S$ (Fig.3).

One of them has the form ($\ref{rotE}$) . If we specify the curve of
integration $s$ as the border of the control area $S$ and denote the tangent
projection \bigskip as $E_{t}=\overrightarrow{E}\cdot\overrightarrow{t}$ we
have
\begin{equation} 
{\displaystyle\int\limits_{s}}
E_{t}ds=0. \label{ped} 
\end{equation}

The next one is the first equation of Navier- Stokes ($\ref{NS-1}$)
:
\begin{equation}
W\frac{\partial W}{\partial y}=G_{r}\left(  T+1\right)  -\frac{p_{\infty} 
}{\rho_{\infty}W_{o}^{2}}\frac{\partial p}{\partial y}+\left(  \frac
{\partial^{2}W}{\partial y^{2}}+\frac{\partial^{2}W}{\partial x^{2}}\right)
\label{Be} 
\end{equation}
The excess pressure is the case of natural heat convection problem to be
considered is small and proportional to the excess of the temperature: 
\begin{equation}
p=\left(  \frac{dp}{dT}\right)  _{\rho=\rho_{\infty}}\left(  T-T_{\infty
}\right)  =\rho_{\infty}R^{\prime}\left(  T-T_{\infty}\right)  . \label{p} 
\end{equation}
In the case of the gas one can use the ideal gas equation of state in
dimensional form $\widetilde{p}=\rho R^{\prime}\left[  T_{\infty}+\left(
T-T_{\infty}\right)  \right]  =p_{0}+p$ $\thickapprox\rho_{\infty}R^{\prime
}\left[  T_{\infty}+\left(  T-T_{\infty}\right)  \right]  $ where $p_{0} 
=\rho_{\infty}R^{\prime}T_{\infty}$ \ and $R\prime=R/\mu$ ($\mu$ is a molar
mass) is gas constant hence $p=\rho_{\infty}R^{\prime}\left(  T-T_{\infty
}\right)  .$
Going to the nontraditional form, plugging ($\ref{p}$) to ($\ref{Be}$) yields 
\begin{equation}
W\frac{\partial W}{\partial y}=G_{r}\left(  T+1\right)  -H\frac{\partial
T}{\partial y}+\left(  \frac{\partial^{2}W}{\partial y^{2}}+\frac{\partial
^{2}W}{\partial x^{2}}\right)  \label{NS1} 
\end{equation}
where we introduced the new nontraditional parameter
\begin{equation}
H=\frac{\left(  T_{w}-T_{\infty}\right)  R^{\prime}}{W_{o}^{2}}=\frac
{\Phi^{\frac{1}{3}}R^{\prime}}{\left(  gb\nu^{\prime}\right)  ^{\frac{2}{3}} 
}G_{r}^{\frac{2}{3}}, \label{H} 
\end{equation}
where $T_{w}-T_{\infty}=\Phi,W_{o}$ and $G_{r}$ are defined by ($\ref{W0}$)
and ($\ref{Gr}$) \ . The viscosity coefficient $\nu^{\prime}$ here is dimensional.

The next boundary condition is connected with the conservation of energy in a
control volume $V$ (area S with unit width see Fig.3) arises from FK equation
($\ref{FK}$) by integration over the volume.\bigskip 
\begin{equation}
\Pr 
{\displaystyle\int\limits_{V}}
\left(  W\frac{\partial T}{\partial y}\right)  dV= 
{\displaystyle\int\limits_{V}}
\left(  \frac{\partial^{2}T}{\partial y^{2}}+\frac{\partial^{2}T}{\partial
x^{2}}\right)  dV= 
{\displaystyle\int\limits_{S}}
\left(  \operatorname{grad}T\right)  \overrightarrow{n}dS \label{En} 
\end{equation}
Let us now formulate the rest four boundary conditions on the base of the
conservation laws.

According to mass conservation law ($\ref{mass}$) the mass flux don't depend
on $y$. If \ we recall that the mean density is approximately constant, then
$\left[  \frac{\partial}{\partial y} 
{\displaystyle\int\limits_{0}^{1}}
\rho_{\infty}Wdx\right]_{y\rightarrow0}=0,$ or
$\left[  \frac{\partial}{\partial y} 
{\displaystyle\int\limits_{0}^{1}}
\left(  \gamma x+\alpha\left( y\right)  x^{2}+\beta\left(  y\right)
x^{3}\right)  dx\right]  _{y\rightarrow0}=\allowbreak0.$Integration and
differentiation yield the third boundary condition: 
\begin{equation}
\left[  \frac{1}{3}\frac{\partial\alpha\left(  y\right)  }{\partial y} 
+\frac{1}{4}\frac{\partial\beta\left(  y\right)  }{\partial y}\right]
_{y\rightarrow0}=0. \label{war-3} 
\end{equation}
Plugging the explicit expressions for $\alpha\left(  y\right)  $ ($\ref{C1C2} 
$) and $\beta\left(  y\right)  $ ($\ref{BETA}$) into ($\ref{war-3}$) yields
It gives a link between the constant of integration $C_{1}$\ and the boundary
value of $\frac{\partial C\left(  y\right)  }{\partial y}=C^{\prime}(y)$ at
the point of $y=0$ 
\begin{equation}
C_{1}=\frac{G_{r}}{8}C^{\prime}(0). \label{C-1} 
\end{equation}
Let us apply the conservations laws in integral form to the control volume $V$
(see Fig.3) based on the interval $y\in\left[  0,\varepsilon\right]  $ and
$x\in\left[  0,1\right]  $. According to our main assumption about
two-dimensionality of the stream we neglect a dependence of variables on $z$ coordinate. 
\begin{figure}
[ptb]
\begin{center}
\includegraphics[
natheight=2.905800in,
natwidth=5.137800in,
height=1.983in,
width=3.4921in
] 
{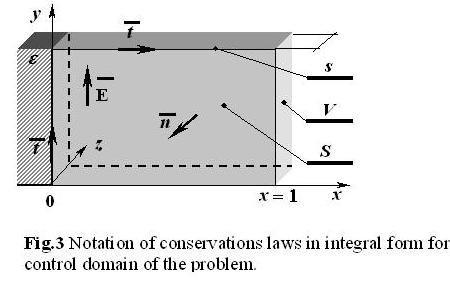} 
\end{center}
\end{figure}
Going to the momentum conservation let us substitute the components of the
vector $\overrightarrow{E}$ :
into ($\ref{NS-a}$) or plugging\ \bigskip$E_{y\text{ }}=\frac{\partial
p}{\partial y}=W\frac{\partial W}{\partial y}-G_{r}\left(  T+1\right)
-\left(  \frac{\partial^{2}W}{\partial y^{2}}+\frac{\partial^{2}W}{\partial
x^{2}}\right)  $ into ($\ref{ped}$) results in\ (see Fig.2): 
\begin{equation} 
{\displaystyle\oint\limits_{s}}
E_{t}ds= 
{\displaystyle\int\limits_{0}^{\varepsilon}}
E_{y}|_{x=0}dy=0, \label{ped1} 
\end{equation}
because the rest parts of the integral are zero according to assumption that
velocity component $W_{x}=0$ is zero $E_{x}=0.$

The left side of the equation ($\ref{ped1}$) according to ($\ref{x'}$) and the
fact that the velocity $W$ vanishes on the vertical plate surface has the form :
$
{\displaystyle\int\limits_{0}^{\varepsilon}}
E_{y}|_{x=0}dy=$ $G_{r}\varepsilon+ 
{\displaystyle\int\limits_{0}^{\varepsilon}}
\frac{\partial^{2}W}{\partial x^{2}}|_{x=0}dy=G_{r}\varepsilon+ 
 {\displaystyle\int\limits_{0}^{\varepsilon}}
 \frac{\partial^{2}\left(  \alpha\left(  y\right)  x^{2}\right)  }{\partial
x^{2}}|_{x=0}dy=G_{r}\varepsilon+2 
{\displaystyle\int\limits_{0}^{\varepsilon}}
\alpha\left(  y\right)  dy=0,$
therefore
$G_{r}=-\lim_{\varepsilon\rightarrow0}\frac{2\nu}{\varepsilon} 
{\displaystyle\int\limits_{0}^{\varepsilon}}
\alpha\left(  y\right)  dy=-2\alpha\left(  0\right)  .$

After substitutions of $\alpha(y)$ ($\ref{C1C2}$) and $C_{2}$ from
($\ref{w-I}$) one have two relations.
First one links the integration constant $C_{2}$ with parameters of the
problem  
\begin{equation}
\ C_{2}=-\frac{G_{r}}{2}, \label{C2} 
\end{equation}
while the second one gives the boundary value of the basic function $C\left(
0\right)  $ 
\begin{equation}
C\left(  0\right)  =\frac{12}{G_{r}}\gamma-4. \label{C(0)} 
\end{equation}
Plugging the result into ($\ref{war-2}$) gives: 
\begin{equation}
C^{\prime\prime}(0)=\allowbreak\frac{144}{G_{r}}\gamma-24. \label{C bis} 
\end{equation}
The momentum conservation in differential form is represented by ($\ref{NS1}$)
that we use as a direct source for boundary condition formulation. $\bigskip
$To determine the fifth boundary condition let us evaluate the mean value of
the momentum component balance (shown\ in above equations) that yields the equation:
${\displaystyle\int\limits_{0}^{1}}
\left(
\begin{array}
[c]{c} 
\left(  \gamma x+\alpha\left(  y\right)  x^{2}+\beta\left(  y\right)
\ x^{3}\right)  \left(  \frac{\partial\beta\left(  y\right)  }{\partial
y}x^{3}+\frac{\partial\alpha\left(  y\right)  }{\partial y}x^{2}\right)  -\\
G_{r}\left(  \left(  C\left(  y\right)  x+B\left(  y\right)  x^{3}\right)
+1\right)  +H\left(  \frac{\partial B\left(  y\right)  }{\partial y} 
x^{3}+\frac{\partial C\left(  y\right)  }{\partial y}x\right) \\
-\left(  \frac{\partial^{2}\beta\left(  y\right)  }{\partial y\partial y} 
x^{3}+\frac{\partial^{2}\alpha\left(  y\right)  }{\partial y\partial y} 
x^{2}\right)  -\left(  6\beta\left(  y\right)  x+2\alpha\left(  y\right)
\right)
\end{array}
\right)  dx=0,\allowbreak$

which gives the following at the point $y=0\bigskip\allowbreak$ 
\begin{equation}
-\frac{1}{24}HC^{\prime\prime\prime}(0)\allowbreak+\left(  \frac{1}{2} 
H-\frac{1}{1680}G_{r}^{2}+\frac{13}{3360}G_{r}\gamma\right)  C^{\prime
}(0)+\left(  12\gamma-2G_{r}\right)  \allowbreak=0. \label{war-4} 
\end{equation}
$\bigskip$The last (sixth) boundary condition we found from the energy
conservation equation ($\ref{En}$) for the control volume $V$ shown in the Fig.3 
\begin{equation}
\frac{\Pr}{\varepsilon} 
{\displaystyle\int\limits_{V}}
\left(  W\frac{\partial T}{\partial y}\right)  dV=\frac{1}{\varepsilon} 
{\displaystyle\int\limits_{S}}
\operatorname{grad}T\overrightarrow{n}dS. \label{En-1} 
\end{equation}
Taking into account that integral with respect to $z$ gives the factor $1$, yields
$\frac{\Pr}{\varepsilon} 
{\displaystyle\int\limits_{0}^{\varepsilon}}
\left(
{\displaystyle\int\limits_{0}^{1}}
W\frac{\partial T}{\partial y}dx\right)  dy=\frac{1}{\varepsilon} 
{\displaystyle\int\limits_{0}^{1}}
\left(  -\frac{\partial T}{\partial y}|_{y=0}+\frac{\partial T}{\partial
y}|_{y=\varepsilon}\right)  dx-\frac{1}{\varepsilon} 
{\displaystyle\int\limits_{0}^{\varepsilon}}
\frac{\partial T}{\partial x}|_{x=o}dy$ \ \ \ \ \

After substitutions $W$ and $T$ ($\ref{twyx}$), the left side of ($\ref{En-1} 
$) gives:

$\allowbreak\bigskip\Pr\left(  \frac{\gamma}{70}-\frac{G_{r}}{504}\right)
\frac{\partial^{3}C\left(  y\right)  }{\partial y\partial y\partial
y}+\allowbreak\allowbreak\Pr\left(  \frac{G_{r}}{120}-\frac{\gamma} 
{15}\right)  \frac{\partial C\left(  y\right)  }{\partial y},$while the right
side after transition $\varepsilon\rightarrow0$ tends to:$\allowbreak\left[
\frac{1}{4}\frac{\partial^{2}B\left(  y\right)  }{\partial y\partial y} 
+\frac{1}{2}\frac{\partial^{2}C\left(  y\right)  }{\partial y\partial
y}\right]  _{y=0}-C\left(  0\right)  .$

Plugging the expression for $B$ ($\ref{BC}$) and next using the
equation ($\ref{C}$) for the basic function $C(y)$ gives:

$\left[  \allowbreak\frac{1}{2}\frac{\partial^{2}C\left(  y\right)  }{\partial
y\partial y}-\frac{1}{24}\frac{\partial^{4}C\left(  y\right)  }{\partial
y\partial y\partial y\partial y}\right]  _{y=0}-C\left(  0\right)
=\allowbreak-\frac{G_{r}\text{ }\Pr}{8}C^{\prime}(0)+\frac{60}{G_{r}} 
\gamma-8.$

The complete equation ($\ref{En-1}$) gives the last boundary relation: 
\begin{equation}
\Pr\left(  \frac{\gamma}{70}-\frac{G_{r}}{504}\right)  C^{\prime\prime\prime
}(0)+\allowbreak\allowbreak\Pr\left(  \frac{G_{r}}{120}-\frac{\gamma} 
{15}\right)  C^{\prime}(0)=\allowbreak-\frac{G_{r}\Pr\text{ }}{8}C^{\prime
}(0)+\frac{60}{G_{r}}\gamma-8. \label{war-VI} 
\end{equation}

\section{The explicit form of boundary conditions for C(y)}

Let us solve the conditions ($\ref{war-4}$) and ($\ref{war-VI}$) with respect
to the first and third derivatives of $C\left(  y\right)  $ at \ $y=0.$

The relation ($\ref{war-4}$) \ yields at $y=0:-\frac{1}{24}HC^{\prime
\prime\prime}(0)\allowbreak+\left(  \frac{1}{2}H-\frac{1}{1680}G_{r}^{2} 
+\frac{13}{3360}G_{r}\gamma\right)  C^{\prime}(0)+12\gamma-2G_{r} 
\allowbreak=0.$

Plugging $C^{\prime\prime\prime}(0)\allowbreak=\left(  12-\frac{1}{70H} 
G_{r}^{2}+\frac{13}{140H}G_{r}\gamma\right)  C^{\prime}(0)+\frac{288}{H} 
\gamma-\frac{48}{H}G_{r}\allowbreak$ into ($\ref{war-VI}$) $\allowbreak
\allowbreak\allowbreak\allowbreak$ gives the equation for $\left[
\frac{\partial C\left(  y\right)  }{\partial y}\right]  _{y=0}=M$,

$\bigskip\allowbreak\allowbreak\allowbreak\allowbreak\left(  \frac{\gamma} 
{70}-\frac{G_{r}}{504}\right)  \left[  \left(  12-\frac{G_{r}^{2}}{70H} 
+\frac{13G_{r}\gamma}{140H}\right)  C^{\prime}(0)+\frac{288\gamma}{H} 
-\frac{48G_{r}}{H}\allowbreak\right]  +\left(  \frac{G_{r}}{120}-\frac{\gamma
}{15}\right)  M+\frac{G_{r}\text{ }M}{8}-\frac{60\gamma}{\Pr G_{r}}+\frac
{8}{\Pr}=0$

Solving this linear equation yields: 
\begin{equation}
\bigskip M=\frac{-\frac{8}{\Pr}+\frac{24}{H}\left(  2G_{r}-12\gamma\right)
\left(  \frac{1}{70}\gamma-\frac{1}{504}G_{r}\right)  +\frac{60}{\Pr G_{r} 
}\gamma}{-\frac{1}{15}\gamma+\frac{2}{15}G_{r}+\frac{24}{H}\left(  \frac
{1}{70}\gamma-\frac{1}{504}G_{r}\right)  \left(  \frac{1}{2}H-\frac{1} 
{1680}G_{r}^{2}+\frac{13}{3360}G_{r}\gamma\right)  }. \label{M} 
\end{equation}
where $H$ is defined by ($\ref{H}$)
The result allows to evaluate the third derivative at $y=0$. Let us list the
boundary conditions for $C(y):$ 
\begin{align}
\bigskip C\left(  0\right)   &  =\left(  \frac{12}{G_{r}}\gamma-4\right)
,C^{\prime}\left(  0\right)  =M,\bigskip C^{\prime\prime}\left(  0\right)
=\frac{144}{G_{r}}\gamma-24,\label{KondC}\\
C^{\prime\prime\prime}\left(  0\right)   &  =\left(  12-\frac{1}{70H}G_{r} 
^{2}+\frac{13}{140H}G_{r}\gamma\right)  M+\frac{288}{H}\gamma-\frac{48} 
{H}G_{r}\text{\allowbreak.}\nonumber
\end{align}
\section{The solution of the boundary problem for C(y)}
Let us rewrite the equation ($\ref{C}$) substituting the expressions for
$C_{1}(\ref{C-1})$ and $C_{2}$ ($\ref{C2}$) and introducing the Rayleigh
number $R_{a}$ = $G_{r}\Pr$ . We have 
\begin{equation}
\frac{\partial^{4}C\left(  y\right)  }{\partial y\partial y\partial y\partial
y}+3R_{a}\left(  \frac{M}{4}y-1\right)  \frac{\partial C\left(  y\right)
}{\partial y}=0. \label{CC} 
\end{equation}
As it is seen from estimation of $M$ $(Gr)$ one can neglect the term $\frac
{M}{4}y<<1$ in comparison with unit within the interval $y\in\lbrack0,10]$ if
the Grashof number $Gr>500$ . Such approximation yields the equation with
constant coefficients with solution as linear combination of exponents:
\begin{equation}
C(y)=\sum_{i=0}^{3}A_{i}exp[k_{i}y], \label{sol} 
\end{equation}
where $k_{i}$ are roots of the equation
\begin{equation}
k^{4}-3R_{a}k=0. \label{keq} 
\end{equation}
It means that
\begin{equation}
k_{0}=0,k_{1}=\sqrt[3]{3R_{a}}\equiv s,k_{2,3}=(-\frac{1}{2}\pm i\frac
{\sqrt{3}}{2})s. \label{ki} 
\end{equation}
Plugging (\ref{ki}) into (\ref{sol}), taking into account reality of $C(y)$,
results in:
\begin{equation}
C(y)=A_{0}+A_{1}\exp[sy]+\exp[-\frac{sy}{2}](B_{1}\cos[\frac{\sqrt{3}} 
{2}sy]+B_{2}\sin[\frac{\sqrt{3}}{2}sy]). \label{sol} 
\end{equation}
The second term $A_{1}\exp[sy]$ in the above expression exponentially grows
for positive $s$, therefore the value of $A_{1}$\ should be analyze in details
especially for large values of $s$. \

Boundary conditions (\ref{KondC}) subsequently give 
\begin{equation}
C(0)=A_{0}+A_{1}+B_{1}=\left(  \frac{12}{G_{r}}\gamma-4\right)  , \label{C0} 
\end{equation} 
\begin{equation}
C^{\prime}(0)=s\left(  A_{1}-B_{1}\frac{1}{2}+B_{2}\frac{\sqrt{3}}{2}\right)
=M, \label{C1} 
\end{equation} 
\begin{equation}
C^{\prime\prime}\left(  0\right)  =s^{2}(A_{1}-\frac{B_{1}}{2}-\frac{\sqrt
{3}B_{2}}{2})=\frac{144}{G_{r}}\gamma-24, \label{C2} 
\end{equation}
\begin{equation}
C^{\prime\prime\prime}\left(  0\right)  =s^{3}(A_{1}+B_{1})=\left(
12-\frac{1}{70H}G_{r}^{2}+\frac{13}{140H}G_{r}\gamma\right)  M+\frac{288} 
{H}\gamma-\frac{48}{H}G_{r}\text{\allowbreak}, \label{C3} 
\end{equation}
Solving the system of above equations with respect to $A_{i}$ and $B_{i}$
yields
\begin{equation}
B_{2}=\frac{1}{3}\sqrt{3}\left(  \frac{M}{s}+\frac{24}{s^{2}}-\frac{144} 
{G_{r}s^{2}}\gamma\right)  \label{B2} 
\end{equation}
\begin{equation}
B_{1}=\allowbreak8\frac{M}{s^{3}}-\frac{1}{3}\frac{M}{s}+\frac{8}{s^{2}} 
+\frac{192}{Hs^{3}}\gamma-\frac{32}{Hs^{3}}G_{r}-\frac{48}{s^{2}}\frac{\gamma
}{G_{r}}-\frac{1}{105H}\allowbreak\frac{M}{s^{3}}G_{r}^{2}+\frac{13} 
{210H}\frac{M}{s^{3}}\gamma G_{r} \label{B1} 
\end{equation}
\begin{equation}
A_{1}=\frac{1}{3}\frac{M}{s}+4\frac{M}{s^{3}}-\frac{8}{s^{2}}+\frac{96} 
{Hs^{3}}\gamma-\frac{16}{Hs^{3}}G_{r}+\frac{48}{s^{2}}\frac{\gamma}{G_{r} 
}-\frac{1}{210H}\allowbreak\frac{M}{s^{3}}G_{r}^{2}+\frac{13}{420H}\frac
{M}{s^{3}}\gamma G_{r}, \label{A1} 
\end{equation} 
\begin{equation}
A_{0}=-\frac{\left(  12-\frac{1}{70H}G_{r}^{2}+\frac{13}{140H}G_{r} 
\gamma\right)  M+\frac{288}{H}\gamma-\frac{48}{H}G_{r}}{s^{3}}+\left(
\frac{12}{G_{r}}\gamma-4\right)  . \label{Azero} 
\end{equation}
The parameter $\gamma$ in all above expressions is still not defined$.$The
physical meaning of it relates to the velocity profile, namely it is the angle
of inclination on the plate ($\ref{twyx}$)$.$ In the literature the profiles
are given either directly from experimental data \cite{4,6} or from
theoretical models \cite{2}. \ For example in the book \cite{2} the velocity
dimensionless counterpart $U=$ $\frac{d\xi}{d\eta}$ (in our paper is denoted
as $W$) in appropriate units is plotted against the function $\eta$ for
different Prandtl numbers. The velocity $U=u\frac{x}{2\nu\sqrt[2]{Gr_{x}}}$ is
the function of $y$ which is normal directed to the plate and $x$ is along the
plate (opposite to ours notations). Numerically differentiating the expression
for $U$ with respect to $y$ on the level $x=l$ and taking into account the
expression ($\ref{twyx}$) we derive the link between constant $\gamma$ and the
derivative $\frac{dU}{d\eta}|_{\eta=0}=\zeta:$
\begin{equation}
\gamma=\sqrt{2}G_{r}^{3/4}\zeta.\label{gamalink} 
\end{equation}
The alternative approach to the solution of the equation ($\ref{CC})$ is the
power series expansion. To investigate the parameter values case of order one
we use the Frobenius method. According to the method let us solve the equation
($\ref{CC}$) as a power series of $y$:\bigskip 
\begin{equation}
C(y)=a_{0}+a_{1}y+a_{2}y^{2}+a_{3}y^{3}+a_{4}y^{4}+a_{5}y^{5}+... \label{C(Y)} 
\end{equation}
$\bigskip$The equations ($\ref{CC})$ and ($\ref{C(Y)})$ yield the links:
$a_{4}$ $=\frac{1}{8}a_{1}R_{a}$, $a_{5}=$ $\frac{1}{20}a_{2}R_{a}-\frac
{1}{160}Ma_{1}R_{a}$, $a_{6}=$ $\frac{1}{40}a_{3}R_{a}-\frac{1}{240} 
Ma_{2}R_{a}.$ The boundary conditions give:
\begin{equation}
a_{0}=C\left(  0\right)  =\frac{12}{G_{r}}\gamma-4, \label{a-0} 
\end{equation}
\begin{equation}
a_{1}=\frac{\partial C\left(  y\right)  }{\partial y}|_{y=0}=M, \label{a1} 
\end{equation}
\begin{align}
a_{2}  &  =\frac{1}{2}\frac{\partial^{2}C\left(  y\right)  }{\partial
y\partial y}|_{y=0}=\frac{72}{G_{r}}\gamma-12,\label{areszta}\\
a_{3}  &  =\frac{1}{6}\frac{\partial^{3}C\left(  y\right)  }{\partial
y\partial y\partial y}|_{y=0}=\left(  2-\frac{1}{420H}G_{r}^{2}+\frac
{13}{840H}G_{r}\gamma\right)  M+\frac{48}{H}\gamma-\frac{8}{H}G_{r} 
\allowbreak,\nonumber\\
a_{4}  &  =\frac{1}{8}MR_{a},a_{5}=\frac{1}{20}\left(  \frac{72}{G_{r}} 
\gamma-12\right)  R_{a}-\frac{1}{160}M^{2}R_{a}\ ,\nonumber\\
a_{6}  &  =\allowbreak\allowbreak\frac{R_{a}}{10}\left(  1-\frac{G_{r}^{2} 
}{16\,80H}-\frac{3}{G_{r}}\gamma+\frac{13G_{r}}{33\,60H}\gamma\right)
\allowbreak M+\frac{R_{a}}{5}\left(  \frac{6}{H}\gamma-\frac{G_{r}}{H}\right)
.\nonumber
\end{align}

\section{\bigskip The example of air}
\subsection{\bigskip A solution of first type}

For representation of results we choose air as a fluid with typical parameters
$T_{av}$ = $20^{o}C,p=760$ $Tr,Pr=0.7$ and assume $\Phi=10K$. The physical
table values in such conditions are following: $a=\allowbreak2.\,\allowbreak
142\times10^{-5}\frac{m^{2}}{s},\bigskip\nu=\allowbreak1.\,\allowbreak
506\times10^{-5}\frac{m^{2}}{s},R=287.06$ $\frac{J}{kgK},b=\allowbreak
3.\,\allowbreak43\,0\times10^{-3}\frac{1}{K}.$For such values the parameter
$H$ ($\ref{H})$ is:
\begin{equation}
H=9.\,\allowbreak606\,4\times10^{6}G_{r}^{\frac{2}{3}}.\label{H1} 
\end{equation}
In the paper \ \cite{2}. an expression of vertical velocity $u(x,y)$ is
derived by the method of similarity solution. For the example of Prandtl
Number $Pr=0.7$ (air) the coefficient is approximately $\zeta=3/5.$Hence the
parameter $\gamma$ ($\ref{gamalink}$) is expressed as: 
\begin{equation}
\gamma=\frac{3\sqrt{2}}{5}G_{r}^{3/4}\label{gamma} 
\end{equation}
Our solution depends on the parameter $M$ via equation ($\ref{CC}$)
coefficient. The condition in which the coefficient is approximately constant
therefore needs the parameter $M$ estimation. For the estimation let us
express the $M$ ($\ref{M}$) as a function of the only parameter $G_{r}$ with
the rest physical parameter values account.\bigskip 
\begin{equation}
\allowbreak M=-\frac{1.\,\allowbreak35\times10^{11}\sqrt[4]{G_{r} 
}-37634.G_{r}^{\frac{4}{3}}+1.\,\allowbreak045\times10^{5}G_{r}^{\frac
{13}{12}}+3360G_{r}^{\frac{19}{12}}-8.\,\allowbreak59\times10^{11} 
}{1.\,\allowbreak05\times10^{9}G_{r}+1.\,\allowbreak3\times10^{9} 
G_{r}^{\frac{5}{4}}-11.\,\allowbreak6 G_{r}^{\frac{7}{3}}+33.\,\allowbreak
7G_{r}^{\frac{25}{12}}+G_{r}^{\frac{31}{12}}}\label{M(G)} 
\end{equation}
 \begin{center}
\fbox{\includegraphics[
natheight=1.420000in,
natwidth=3.541400in,
height=1.42in,
width=3.5414in
] 
{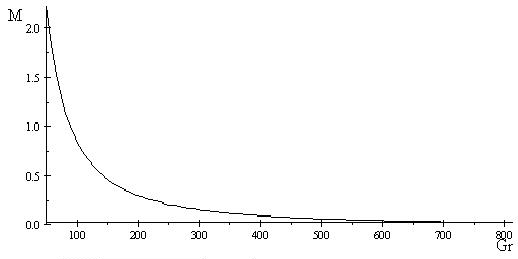} 
}\\
Fig.4 The dependence of the parameter $M$ against Grashof number $G_{r}$ 
\label{Fig4} 
\end{center}
 On the base of the result (see Fig.4) we choose the Grashof number range
$M/4<0.1$ for example evaluations of velocity and temperature profiles as a
function of $x$ and $y.$It means that approximately $G_{r}$ $>200.$

For \ the given values of local Grashof number equal to $G_{r}$ $=200$, $500$
and $1000$ the evaluation of the parameter $\gamma$ by the expression
($\ref{gamalink}$) gives: \bigskip$\gamma=45.\,\allowbreak127$,
$89.\,\allowbreak721$ and $\allowbreak150.\,\allowbreak89.$Next the
calculation of the solution parameters yields the table:

$s=\sqrt[3]{3\cdot G_{r}\cdot0.7}=7.\,\allowbreak488\,9(200)\allowbreak
,=\allowbreak10.\,\allowbreak164(500),=12.\,\allowbreak806(1000);\allowbreak$

$M=\allowbreak\allowbreak0.297\,05(200),$ $6.\,\allowbreak158\,9\times
10^{-2}(500),1.\,\allowbreak200\,4\times10^{-2}(1000);$

$A_{0}=-1.\,\allowbreak300\,9(200),-1.\,\allowbreak847\,5(500),\allowbreak
-2.\,\allowbreak189\,4(1000);$

$\allowbreak A_{1}=6.\,\allowbreak652\,3\times10^{-2}(200),$ $8.\,\allowbreak
191\,1\times10^{-3}(500),$ $-4.\,\allowbreak281\,5\times10^{-3}(1000);$

\allowbreak$B_{1}=$ $-5.\,\allowbreak803\,6\times10^{-2}(200),$
$-7.\,\allowbreak487\,3\times10^{-3}(500),$ $4.\,\allowbreak350\,1\times
10^{-3}(1000);$

$B_{2}=-6.\,\allowbreak452\,2\times10^{-2}(200),$ $-6.\,\allowbreak
786\times10^{-3}(500),8.\,\allowbreak536\,5\times10^{-3}(1000).\bigskip$

Analysis of the dependence of the coefficient $A_{1\text{ }}$on Grashof number
shows its change of sign (see the values for $A_{1}$ for $G_{r}$ $=500$ and
$1000$). Plugging the expression for $M$ ($\ref{M(G)}$) into the formula for
$A_{1}$ ($\ref{A1}$) and its numerical evaluation allow to plot the dependence
$A_{1}(G_{r}).$The corespondent plot for the case of air is given below
(Fig.5) and allow to determine a critical value of the number $G_{r} 
=G_{r,cr.}$for which the characteristic linear dimension of the problem has
the concrete value $l=\left(  \frac{G_{r,cr}^{2}\cdot\nu^{2}}{g\cdot
b\cdot\Phi}\right)  ^{1/3}$. 
 \begin{center}
\fbox{\includegraphics[
natheight=1.863700in,
natwidth=4.330100in,
height=1.8637in,
width=4.3301in
] 
{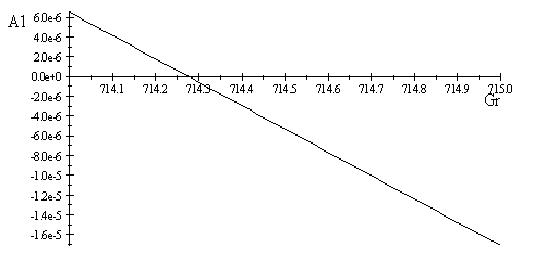} 
}\\
Fig.5 The dependence of the $A_{1}$ coefficient as the function of $G_{r}$ in
the vicinity of the critical value $G_{r,cr}$.
\label{Fig.5} 
\end{center}
For $A_{1}\thickapprox0$, form the plot we estimate $G_{r}=714.275,$\bigskip
and next calculations give $\gamma=150.\,\allowbreak89,s=\allowbreak
11.\,\allowbreak447,M=2.\,\allowbreak916\,2\times10^{-2},$ 
\begin{equation}
A_{0}=-2.\,\allowbreak030\,7,B_{2}=3.\,\allowbreak074\,8\times10^{-3} 
,B_{1}=2.\,\allowbreak333\,2\times10^{-4}. \label{par} 
\end{equation}

Plugging the all the values for four cases ($G_{r}=200,500,714.275$ and
$1000$) into the solution ($\ref{sol}$) we plot the resulting curves (Fig.6).
  \begin{center}
\fbox{\includegraphics[
natheight=1.898300in,
natwidth=4.440800in,
height=1.8983in,
width=4.4408in
] 
{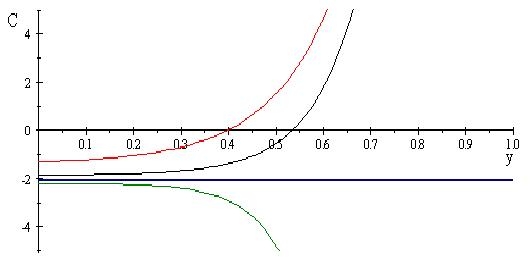} 
}\\
Fig.6 The dependence of the basic function C on y for Gr=200 (red), =500
(black), 714.275 (blue) and =1000 (green).
\end{center}
As we announced in the introduction this study is focused on the problem of
the starting point of the stationary flow in the vicinity of the point $x=0$,
$y=0$.

For further illustration of the velocity and temperature profiles we choose
the critical value of Grashof number $G_{r,cr}=$ $714.275$. Pugging the
parameters values ($\ref{par}$) into formulas for the velocity and temperature
($\ref{twyx}$) we obtain the profiles of velocity $W\left(  0,x\right)  $ $=$
$\gamma x+\left(  -\frac{G_{r,cr}}{2}\right)  x^{2}+\left(  -\frac{G_{r,cr} 
}{6}\left(  \frac{12}{G_{r,cr}}\gamma-4\right)  \right)  x^{3}$and $W\left(
1,x\right)  =\gamma x+\left(  \frac{G_{r,cr}}{8}My-\frac{G_{r,cr}}{2}\right)
x^{2}+\left(  -\frac{G_{r,cr}}{6}C\left(  1\right)  \right)  x^{3}$ on the
Fig.7 and temperature obtained in similar way on Fig.8. 
 \begin{center}
\fbox{\includegraphics[
natheight=1.365500in,
natwidth=4.299000in,
height=1.3655in,
width=4.299in
] 
{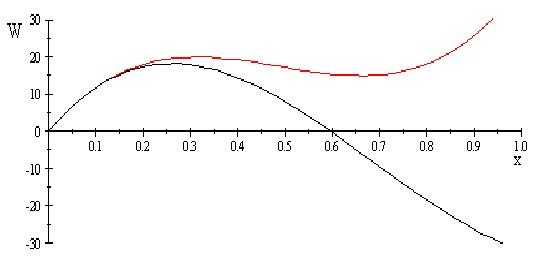} 
}\\
Fig.7 Profiles of velocity along x on levels y = 0(black) y=1 (red)
\label{Fig.7} 
\end{center}
 \begin{center}
\fbox{\includegraphics[
natheight=1.407900in,
natwidth=4.357800in,
height=1.4079in,
width=4.3578in
] 
{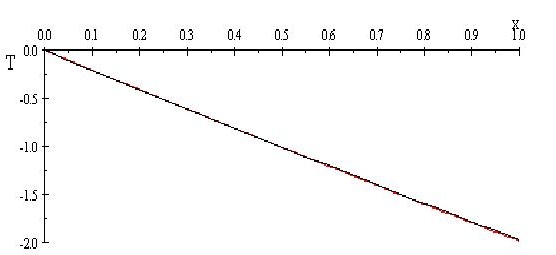} 
}\\
Fig.8 Dependence of temperature on x at the level y=0 (black) and y=1 (red).
\end{center}

\subsection{ The solution of the second type for C(y)}

\bigskip Let us recall the basic relations that define a solution of
Navier-Stokes and Fourier-Kirchhoff equations via the power series in $y$.
Plugging expressions for coefficients $\gamma,\alpha\left(  y\right)
,\beta\left(  y\right)  ,C(y)$ and $B(y)$ in velocity and temperature
expansions ($\ref{twyx}$) one have 

\begin{equation}
W\left(  y,x\right)  =x\gamma+x^{2}\frac{G_{r}}{2}\left(  \frac{1} 
{4}My-1\right)  -x^{3}\frac{G_{r}}{6}C\left(  y\right)  \label{velo} 
\end{equation}
\begin{equation}
T\left(  y,x\right)  =C\left(  y\right)  x-\frac{1}{6}\frac{\partial
^{2}C\left(  y\right)  }{\partial y\partial y}x^{3} \label{temp} 
\end{equation}
where: $C(y)=a_{0}+a_{1}y+a_{2}y^{2}+a_{3}y^{3}+a_{4}y^{4}+a_{5}y^{5}$ with
the coefficients from ($\ref{a0}$), ($\ref{a1}$) and ($\ref{areszta}$).
According to the physical essence of temperature profile the function $C(y)$
should be negative. It means that at least the coefficient $C(1)=a_{0} 
+a_{1}+a_{2}+a_{3}+a_{4}+a_{5}<0.$In such condition of $0<y<1$ the function
$C(y)<0.$ Plugging the expressions for $\gamma,H$ and $M$ to $C(1)$ yields the
complicated function of $G_{r}$. The result is illustrated in Fig.9.
    \begin{center}
\fbox{\includegraphics[
natheight=2.035800in,
natwidth=4.711500in,
height=2.0358in,
width=4.7115in
] 
{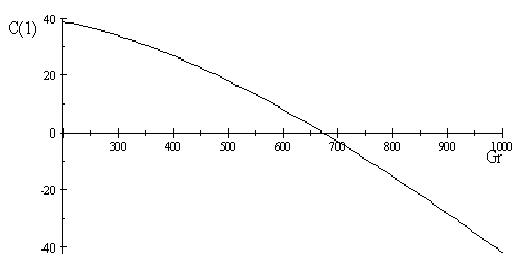} 
}\\
Fig.9 The enlarged fragment of  C(1)  the vicinity of upper critical
Grashoff number.
\label{Fig.9b} 
\end{center}

From the plot Fig.9b it follows that the values of $C(1)$ are negative if
$G_{r}$ $>G_{r,0}$. For \ an illustration we investigate the approximate
solution with the local Grashof number value equal to $G_{r}$ $=676.5$, for
which $C(1)=-0.455\,29$ that lays inside the interval $(0,-1)$ which guarantee
the negative temperature values for all $x\in(0,1).$
For the plotting we need the following general parameters ($\ref{M(G)}$) for $G_{r}$ $=676.5$. The values are: $\gamma
=112.\,\allowbreak56,H=$ $7.\,\allowbreak403\,0\times10^{8},M=\allowbreak
3.\,\allowbreak297\,4\times10^{-2}$and the coefficients ($\ref{a1},\ref{areszta}$) of the polynomial $C(y)$:
$a_{0}=\allowbreak-2.\allowbreak003\,4,a_{1}=M=3.\,\allowbreak
3\times10^{-2},a_{2}=\allowbreak-0.0207,a_{3}=6.\,\allowbreak
594\,8\times10^{-2},a_{4}$ $=\frac{1}{8}MR_{a}=\allowbreak1.\,\allowbreak
951\,9,$
$a_{5}=$ $\allowbreak-0.482\,02.$ Substituting these values into
($\ref{C(Y)}$) gives:
$C(y)=\allowbreak-4.\,\allowbreak560\,5y^{5}+1.\,\allowbreak
822\,6\allowbreak y^{4}+5.\,\allowbreak832\,5\times10^{-2}y^{3} 
-0.182\,32\allowbreak y^{2}+2.\,\allowbreak916\,2\times10^{-2} 
y-2.\,\allowbreak030\,4\allowbreak. $In similar way we derive the expression of
C(y) for the critical Graschoff number of the previous solution $G_{r,cr} 
=714.275$. Below we plot both curves on the Fig.10. 
 \begin{center}
\fbox{\includegraphics[
natheight=2.097200in,
natwidth=4.368200in,
height=2.0972in,
width=4.3682in
] 
{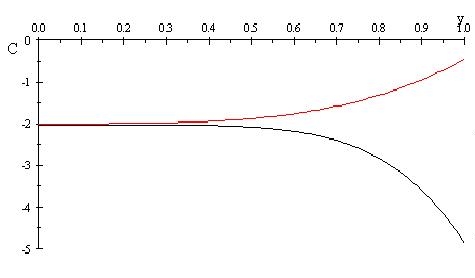} 
}\\
Fig.10 The power series representation for C(y) at Grashoff number Gr= 676,5
(red) and $G_{r,cr}$ = 714.275 (black).
\label{Fig.10} 
\end{center}
The reason of discrepancy between the curves is explained mainly by the
dependence of the coefficient $a_{2}$ on Graschoff number (Fig.11). 
 \begin{center}
\fbox{\includegraphics[
natheight=1.996000in,
natwidth=4.385500in,
height=1.996in,
width=4.3855in
] 
{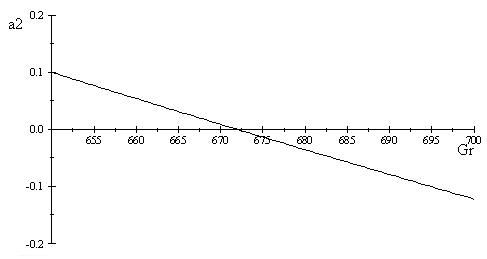} 
}\\
Fig.11 Dependance of a2 on Grashoff number.
\label{Fig.11} 
\end{center}
As one can see the value of $a_{2}$ change the sign at about $G_{r}=672$

Finally let us substitute the function of C(y) for the $G_{r,cr}=714.275$ into
the expressions for the velocity and temperature and plot them at the levels
$y=0$ and $y=0.5$(see Fig.12 and Fig.13). 
\begin{center}
\fbox{\includegraphics[
natheight=2.028000in,
natwidth=4.347400in,
height=2.028in,
width=4.3474in
] 
{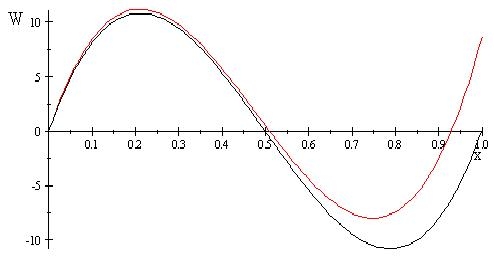} 
}\\
Fig.12 Velocity profiles for the second solution at the y=0 level (black) and
y=0.5 (red).
\label{Fig.12} 
\end{center}
 \begin{center}
\fbox{\includegraphics[
natheight=2.000300in,
natwidth=4.423500in,
height=2.0003in,
width=4.4235in
] 
{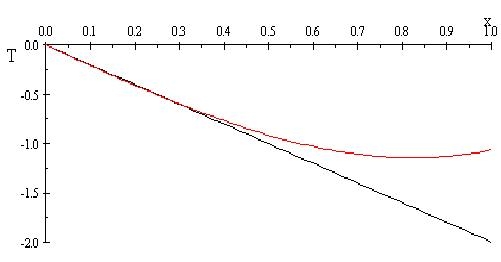} 
}\\
Fig.13 Temperature profiles at  levels y=0 (black) and y=0.5 (red).
\label{Fig.13} 
\end{center}
\section{Analysis of the solution and conclusion}
$\allowbreak$According to the order of power series expansion in $x$ the
expressions for velocity and temperature in the form of ($\ref{velo} 
)$,\ \ \ ($\ref{temp})$ satisfy the N-S and the F-K equations up to the term
$x^{3}$. Hence the solution is valid in a stripe adjacent to the surface. The
equation for the coefficient function $C$ $(\ref{CC})$ may be applied on the
interval $y\in\lbrack0,\infty)$ but the approximate solutions we study here
give satisfactory results in a vicinity of starting point of the flow $y=0$.

The present theory allows to include higher terms of the expansion hence to
obtain results that are valid in more wide range of coordinates.
Our calculatins were performed for the air case that corresponds to the choice
of state equation in ideal gas form and Prandtl number $Pr=0.7$. There is an
obvious possiblility to extend our results for other fluids, for example as in
\cite{5}, \cite{6}.  

\end{document}